\documentclass[11pt,a4paper]{article}
\usepackage{jstyle}

\newcommand{\syd}{%
{\begin{tikzpicture}
\draw (0,0) rectangle (0.2,0.1);
\draw (0.1,0) -- (0.1,0.1);
\end{tikzpicture}}}

\newcommand{\ayd}{%
{\begin{tikzpicture}
\draw (0,0) rectangle (0.1,0.2);
\draw (0,0.1) -- (0.1,0.1);
\end{tikzpicture}}}

\author{Euihun JOUNG}
\author{\quad Karapet MKRTCHYAN}

\affiliation{School of Physics \& Astronomy and Center for Theoretical Physics \\ Seoul National University, Seoul 08826 \rm KOREA}
\affiliation{Gauge, Gravity \& Strings, \ Center for Theoretical Physics of the Universe\\
Institute for Basic Sciences, Daejeon 34047 \rm KOREA}

\emailAdd{euihun.joung@snu.ac.kr}
\emailAdd{karapet@snu.ac.kr}

\title{\centering
\LARGE{Weyl Action of
Two-Column Mixed-Symmetry Field \\[5pt]
and 
Its Factorization Around
 (A)dS Space}}

\abstract{We investigate the four-derivative free 
Weyl action for two-column mixed-symmetry field that makes use of maximal gauge symmetries.
In flat space, the action can be uniquely determined from gauge and Weyl (trace shift) symmetry requirements.
We show that there is a smooth and unique deformation of the flat action to (A)dS
which keeps the same amount of gauge symmetries.
This action admits a factorization into two 
distinct two-derivative actions having 
gauge parameters of different Young diagrams.
Hence, this factorization pattern naturally extends 
that of the Weyl actions of symmetric higher spin fields to mixed-symmetry cases.
The mass-deformation for these actions can be realized preserving one of the gauge symmetries.
Although generically non-unitary, in special dimensions, unitarity is achieved selecting different mass deformations for dS and AdS.
We consider particular examples of our construction such as New Massive Gravity in three dimensions, linearized bigravity in four dimensions and their arbitrary dimensional generalizations.}

\begin{document}

\maketitle

\section{Introduction}

In the realm of the standard field theories, more general fields than symmetric tensors are rarely used. It is partly because we are living in four-dimensional spacetime where any fundamental particle can be described by symmetric tensor(-spinor) fields.
Hence, mixed-symmetry fields --- fields whose indices have more general permutation symmetry than the fully symmetric ones --- would be indispensable only in describing certain physics of higher dimensions, such as the physics of String Theory.
Actually, the infinite tower of massive excitations in String Theory
carry mixed-symmetry representations in general and 
their presence is crucial for the consistency of various stringy dualities.
Moreover, mixed-symmetry fields 
might be useful, although not necessary, even in four dimensions, 
and their use may open a new avenue to unexplored land of physics.

The two-derivative Lagrangian for a massless mixed-symmetry field in flat spacetime is first given for the simplest case by Curtright \cite{Curtright:1980yk} and for generic case by Labastida 
\cite{Labastida:1986gy,Labastida:1986ft,Labastida:1987kw}
and then further investigated  in \cite{Metsaev:1995re,Metsaev:1997nj,Brink:2000ag,Alkalaev:2003hc,deMedeiros:2003qel,
Francia:2005bv,Alkalaev:2006hq,Alkalaev:2005kw,Alkalaev:2006rw, Skvortsov:2008vs, Skvortsov:2008sh, Campoleoni:2008jq, Campoleoni:2009gs,Campoleoni:2009je,
Reshetnyak:2010ga,Alkalaev:2010af,Boulanger:2011qt,Burdik:2011cb,
Campoleoni:2012th,Bekaert:2015fwa,Reshetnyak:2016sng}.
Mixed-symmetry fields can be classified according to the
symmetry property of the index permutations,
hence can be associated with  Young diagrams. 
The Lagrangian has in general a number of gauge symmetries, 
 depending on the shape of the Young diagram associated with the field under consideration.
One of the interesting and non-trivial properties of mixed-symmetry fields is
that their flat space Lagrangian does not admit a smooth 
deformation towards the background with non-vanishing cosmological constant
\cite{Metsaev:1995re,Metsaev:1997nj,Brink:2000ag}:
around (A)dS background,
there is no two-derivative   mixed-symmetry Lagrangian 
respecting all the gauge symmetries  available in flat space.
Instead, one can choose to preserve only one gauge symmetry
and the choice determines the mass squared term to  
a specific value  in units of cosmological constant.

In this paper, we shall focus on the mixed-symmetry field
associated with the  Young diagrams having two columns. 
In a sense, the number of columns plays the role of `spin' hence
we are considering here only `spin two' cases. 
Among generic two-column cases, 
we will analyze  the mixed-symmetry field corresponding to the simple hook Young diagram
$\tiny\yng(2,1)$ \cite{Curtright:1980yk} in great details
since it already contains all the essential features of the more general two-columns cases.
The simple hook field is of particular interest because 
it appears  in the first order formulation of Gravity through the spin connection.
There have been various attempts to describe Gravity by the connection 
or related object
(see for instance the recent discussion \cite{Basile:2015jjd}
and references therein).
We shall also discuss
how these attempts can be understood from the perspective of
the physics of  mixed-symmetry field.

The main target of the current paper is the identification and the analysis 
of the higher-derivative action of mixed-symmetry field having Weyl transformation --- that is, the trace shift --- as its symmetry. In the more familiar case of symmetric spin-two field,
the Weyl action is nothing but the linearization
of the four-derivative  Weyl gravity.
The latter can be decomposed around (A)dS
into the massless and partially-massless modes (see \cite{Deser:2012qg} for
the related discussion).
In the case of symmetric spin-$s$ field, the Weyl action involves $2s$ derivatives 
which can be split into $s$ different  modes each of which
is described by a specific two-derivative action with a certain gauge symmetry
\cite{Joung:2012qy,Metsaev:2014iwa,Nutma:2014pua}:
they are partially-massless spin $s$ modes of depth $0$ to $s-1$\,.
In fact, these are all the known short representations (of isometry group) containing 
the helicity $s$ mode as the highest one.

This pattern suggests that the Weyl action of certain type of field ---
fully symmetric or mixed symmetry ---
could describe all the short representations associated with that field
and each of these short representations can be realized as a two-derivative Lagrangian with a certain gauge symmetry (see \cite{Joung:2012qy} for some related discussions).
In this paper, we examine this idea with 
the mixed-symmetry fields of
two-column Young diagrams.
The Weyl action of the field associated with a Young diagram having $s$ columns ought to
involve $2s$ derivatives. Hence, we expect it to describe $s$ different short modes.
The pattern of  short representations involved might be non-trivial
 and the necessary analysis would be lengthy for generic Young diagrams.
We reserve the general analysis for the future work
and focus here only on the two-column cases where,
irrespectively of the height of each column in the Young diagram,
the Weyl action has four derivatives.
 
Let us provide more details on what has been done in this paper.
We consider the fields having the symmetry of two-column Young diagram,
\be 
\parbox{25pt}{
	\begin{tikzpicture}
	\draw (0,0) rectangle (0.4,2.4);
	\draw (0.4,2.4) -- (0.8,2.4) -- (0.8,0.8) -- (0.4,0.8);
	\node at (0.2,1.2){${\st p}$};
	\node at (0.6,1.6){${\st q}$};
	\end{tikzpicture}}\,.
\ee
There are two short representations in (A)dS described by
two distinct Lagrangians with different gauge symmetries:
the gauge parameters have the index symmetry of Young diagram, either 
\be 
\parbox{45pt}{\begin{tikzpicture}
	\draw (0,0) rectangle (0.4,2);
	\draw (0.4,2) -- (0.8,2) -- (0.8,0.4) -- (0.4,0.4);
	\draw [dotted] (0,0) rectangle (0.4,-0.4);
	\node at (-0.3,1){${\st p-1}$};
	\node at (0.6,1.2){${\st q}$};
	\end{tikzpicture}}\qquad
	{\rm or}\qquad
	\parbox{40pt}{\begin{tikzpicture}
	\draw (0,0) rectangle (0.4,2.4);
	\draw (0.4,2.4) -- (0.8,2.4) -- (0.8,1.2) -- (0.4,1.2);
	\draw [dotted] (0.4,1.2) rectangle (0.8,0.8);
	\node at (0.2,1.2){${\st p}$};
	\node at (1.1,1.8){${\st q-1}$};
	\end{tikzpicture}}\,.
	\label{hk gg sym}
\ee
In the flat limit, the two Lagrangians coincide and enjoy both of the gauge symmetries.
Now considering the Weyl action of this field in flat space,
we find that there is a unique four-derivative action  invariant with respect to both gauge symmetries and a trace shift symmetry with the parameter,
\be 
	\parbox{60pt}{
	\begin{tikzpicture}
	\draw (0,0) rectangle (0.4,2);
	\draw (0.4,2) -- (0.8,2) -- (0.8,0.8) -- (0.4,0.8);
	\draw [dotted] (0,0) rectangle (0.4,-0.4);
	\draw [dotted] (0.4,0.8) rectangle (0.8,0.4);
	\node at (-0.3,1){${\st p-1}$};
	\node at (1.1,1.4){${\st q-1}$};
	\end{tikzpicture}}\,.
\ee
We show that this flat space  Weyl action 
now admits a smooth deformation to the background with non-vanishing cosmological constant,
as opposed to the two-derivative one.
Moreover, this four-derivative action describes
two short-representation modes corresponding to the gauge symmetries
\eqref{hk gg sym}.
In the case where the heights of two columns coincide,
the Weyl action describes,
on top of the usual massless field, 
new degrees of freedom (DoF) having a two-derivative gauge symmetry
similar to the partially-massless spin two field.
Therefore,  our analysis shows 
that the two-column cases precisely fit in the pattern discussed before.

Further, we study two-derivative massive deformations of the Weyl action. In (A)dS there are two distinct mass deformations, which preserve one or the other of the symmetries \eqref{hk gg sym}. We show that the massive action does not admit a smooth (A)dS deformation, similarly to the conventional two-derivative actions. Two different deformations have different spectra around (A)dS background. We observe that the sign of the coefficients in front of the free actions depend on the sign of the cosmological constant. In special dimension \mt{d=p+q+1}, one of the massive deformations is unitary around dS and non-unitary around AdS, while the other deformation follows the opposite pattern --- unitary around AdS and non-unitary around dS. In other words, the unitarity requirement selects one of
the massive deformations, and this choice is different for different signs of cosmological constant.

The plan of the paper is as follows.
We focus first on the  simple hook case in Section \ref{sec:2}.
After a detailed review of the two-derivative systems in Section \ref{sec:2.1}
and \ref{sec:2.2},
we construct and discuss the four-derivative Weyl action in Section \ref{sec:2.3}.
The result of the simple hook is generalized in Section \ref{sec:3}
towards more general two-column cases.
We provide separate discussions on 
 the different height case (Section \ref{sec:3.1}) and
 the equal height case (Section \ref{sec:3.2}).
 Finally, we provide various discussions, in particular, related to New Massive Gravity 
\cite{Bergshoeff:2009hq,Bergshoeff:2012ud,Joung:2012sa,Dalmazi:2012dq},
 in Section \ref{sec:3.3}.

\section{Simple Hook}
\label{sec:2}

Let us first consider the simplest mixed-symmetry field
$\phi_{\m\n,\r}$\,, corresponding to the simple hook Young diagram,
\be
\small{\young(\mu \rho,\nu)}\,.
\ee
The shape of Young diagram dictates the symmetries of the field
under index permutations. The symmetries are 
\be
	\phi_{\m\n,\r}=-\phi_{\n\m,\r}\,,
	\qquad \phi_{\m\n,\r}+\phi_{\n\r,\m}+\phi_{\r\m,\n}=0\,.
\ee
Notice that we work in the base where the anti-symmetry of indices in each column is manifest.
One can equally work with the symmetric base, but for the construction of the 
Lagrangian we find the antisymmetric base more advantageous. 

The Young diagram is fixing not only  how fields transform
under index permutations but also under Lorentz transformations.
Hence, it defines a representation under Lorentz group $SO(1,d-1)$
that the off-shell field carries.

\subsection{Einstein Action in Flat space}
\label{sec:2.1}

In flat spacetime, one can find a proper set of on-shell conditions
--- or, an action principle ---
that makes the field carry the same Young diagram representation 
but now under the little group $SO(d-1)$ or $SO(d-2)$\,, 
depending on whether the field is massive or massless\footnote{Only the representations of compact subgroup $SO(d-2)$ of massless little group $ISO(d-2)$ are relevant for our discussion in this paper.}, respectively.
When given Young diagram representation of Lorentz group does not exist
for the little group, an interesting mechanism may emerge when the propagating modes carry certain Young diagram representations of little group which are different from the Young diagram representation of the Lorentz group that the off-shell field carries. For any mixed-symmetry Young diagram this happens in space-time dimensions lower than certain value.
We shall return to this point later, but work for a while with an arbitrary $d$\,.

In \cite{Curtright:1980yk},
Curtright   constructed the free action for the hook field $\phi_{\m\n,\r}$
which describes massless DoF carrying the hook Young diagram representation
of the little group $SO(d-2)$. Note that such Young diagram exists only when $d$ is greater than four.
By introducing the scalar product,
\be
	\la A\,|\,B\ra=\frac1{m!\,n!}\,\int d^dx\sqrt{|g|}\,A^{\m_1\cdots \m_m,\n_1\cdots \n_n}(x)\,B_{\m_1\cdots \m_m,\n_1\cdots \n_n}(x)\,,
	\label{scalar pr}
\ee
the action for the hook field can be written as
\be
\mathcal S^{\rm\sst flat}_{\rm\sst E}[\phi]=\la \phi\,|\, \cG\,\phi\ra\,.
\label{sh}
\ee
Here the `Einstein tensor' $\mathcal G\,\phi$ is defined 
through the `Ricci tensor' $\mathcal F\,\phi$ as
\be
	\left(\cG\,\phi\right)_{\m\n,\r}=\left(\cF\,\phi\right)_{\m\n,\r}
	-\eta_{\r[\m}\,\left(\cF\,\phi\right)_{\n]\l,}{}^{\l}\,.
\ee
In this paper, the round/square brackets denote full 
symmetrization/anti-symmetrization with weight one,
e.g. $A_{[[\m\n]]}=A_{[\m\n]}=-A_{[\n\m]}$.
The `Ricci tensor' itself is given by
\be
	\left(\cF\,\phi\right)_{\m\n,\r}=\partial^2\,\phi_{\m\n,\r}
	+2\,\partial_{[\mu}\,\partial^\l\, \phi_{\n]\l,\r}
	-\partial_\r\,\partial^\l\,\phi_{\m\n,\l}
	-2\,\partial_{\r}\,\partial_{[\m}\,\phi_{\n]\l\,,}{}^{\l}\,,
\ee
which extends the Fierz-Pauli massless spin two equation to the hook field.
By making use of the generalized Kronecker delta,
\be
	\d_{\m\n\a\b}^{\r\l\g\d}
	=4!\,\d^{\r}_{[\m\phantom{]}}
	\d^{\l\phantom{]}}_{\n\phantom{]}}
	\d^{\g}_{\a\phantom{]}}
	\d^{\d\phantom{]}}_{\b]}\,,
\ee
 the Einstein tensor can be written in a more compact form \cite{Curtright:1980yk} as
\be
	\left(\cG\,\phi\right)_{\m\n,}^{\quad\r}=-\frac12\,
\d_{\m\n\a\b}^{\r\l\g\d}\,\partial^{\a}\,\partial_{\l}\,\phi_{\g\d,}^{\quad\b}\,.
\label{Kr flat}
\ee
The action \eqref{sh} 
has two gauge symmetries,
\be
\d_\e\, \phi_{\m\n,\r}=2\,\partial_{[\m}\,\e_{\n]\r},\qquad
\d_\th\,\phi_{\m\n,\r}=\partial_{\r}\,\th_{\m\n}-\partial_{[\m}\,\th_{\n]\r}\,,
\label{flat gs}
\ee
generated by symmetric and anti-symmetric gauge parameters:
\be
\e_{\m\n}=\e_{\n\m}\,,\qquad
\th_{\m\n}=-\th_{\n\m}\,.
\ee
The above symmetry can be understood diagrammatically as
\be
	\delta_\e\,\young(\phi\phi,\phi)=\young(\e\e,\partial)\,,
	\qquad
		\delta_\th\,\young(\phi\phi,\phi)=\young(\th\partial,\th)\,.
\ee
This gauge symmetry is reducible admitting the gauge-for-gauge symmetry,
\be
	\e_{\m\n}(\xi)=\partial_{(\m}\,\xi_{\n)}\,,\qquad 
	\th_{\m\n}(\xi)=\partial_{[\m}\,\xi_{\n]}\,,\label{gfg}
\ee
which reads  in terms of Young diagram,
\be
	\young(\e\e)=\young(\xi\partial)\,,
	\qquad
	\young(\th,\th)=\young(\xi,\partial)\,.
\ee
The gauge symmetries generated by $\e_{\m\n}(\xi)$ and $\th_{\m\n}(\xi)$ 
have the same form,
\be
	\young(\xi\partial,\partial)\,,
\ee
and therefore sum up to zero for an appropriate choice of the relative coefficient.

The structure of gauge symmetry tells us the number of DoF of this system.
According to the covariant counting (see e.g \cite{Kaparulin:2012px}), we get
\be
 \cS_{\rm\sst E}\quad : \quad  \yng(2,1)_{\sst GL(d)} \ominus 2\left(\yng(2)_{\sst GL(d)} \oplus \yng(1,1)_{\sst GL(d)}\right)\oplus 3\,\yng(1)_{\sst GL(d)}\,,
 \ee
 where we subtract twice the DoF associated to gauge parameters as usual
 and put back three times of the DoF of gauge-for-gauge parameter.
The above can be recast into the counting in the Hamiltonian analysis,
 \be
 	 \yng(2,1)_{\sst SO(d-1)} \ominus \left(\yng(2)_{\sst SO(d-1)} \oplus \yng(1,1)_{\sst SO(d-1)}\right)
 \oplus \yng(1)_{\sst SO(d-1)}\,,
\ee
where we subtract  once the DoF for each of traceless gauge parameters
and put back once that of gauge-for-gauge.
This again can be rearranged into
\be
   \yng(2,1)_{\sst SO(d-1)} \ominus \left(\yng(2)_{\sst SO(d-2)} \oplus \yng(1,1)_{\sst SO(d-2)}
 \oplus \yng(1)_{\sst SO(d-2)}\right) =
   \yng(2,1)_{\sst SO(d-2)}\,.
\ee
which is the DoF of massive hook 
minus those of massless fields $\tiny\yng(2)\,,\ \yng(1,1)$ and $\Box$\,.
Hence, this gives the interpretation of obtaining massless hook 
from the massless limit of massive hook by eliminating other `lower spin' components
the latter involves.
Eventually, we end up with the DoF of the simple hook of the little group $SO(d-2)$\,.
Hence, the DoF of the hook field can be conveniently counted, 
in flat space, through the number of components of the Young diagram representations of the little group.

\subsection{Einstein Action in  (A)dS}
\label{sec:2.2}

Let us move on to the (A)dS background,\footnote{Explicit analysis making use of Stueckelberg fields
can be found in \cite{Zinoviev:2002ye}.}
and consider the analogous Lagrangian to
the flat space one \eqref{sh}.
One can think of the same expression for the Lagrangian as in \eqref{sh} where all flat partial derivatives $\partial_\m$
are replaced by the (A)dS covariant derivatives $\nabla_\m$\,,
but this definition of Lagrangian 
has an ambiguity coming from the commutators of $\nabla_\m$'s
which give terms proportional to (A)dS curvature.
The ambiguous term is hence a non-derivative mass-like term proportional to the cosmological constant,
and it ought to be determined by the invariance of the 
action with respect to either of the gauge symmetries,
\be
\d^{\sst \L}_\e\, \phi_{\m\n,\r}=2\,\nabla_{[\m}\,\e_{\n]\r},\qquad
\d^{\sst \L}_\th\,\phi_{\m\n,\r}=\nabla_{\r}\,\th_{\m\n}-\nabla_{[\m}\,\th_{\n]\r}\,.
\label{hk sym}
\ee
By examining the gauge invariance with a possible mass-like term,
one can realize that it is impossible to preserve both of the gauge symmetries. 
Depending on which symmetry we want to keep, the mass-like term is determined with a different
mass-squared coefficient \cite{Metsaev:1995re,Metsaev:1997nj}.

Let us explictly determine the mass-term or equivalently the Einstein tensor $\cG_{\m\n,\r}$
of the (A)dS Lagrangian defined in the same manner as in \eqref{sh}. 
To proceed, let us first fix our conventions:
\be
[\nabla_\m, \nabla_\n]\,V_{\l}^{\r}=R_{\m\n,\l}{}^{\s}\,V_{\s}^{\r}-
R_{\m\n,\s}{}^{\r}\,V_{\l}^{\s}\,,\qquad
R_{\m\n,\l}{}^{\r}=\frac{2\,\L}{(d-1)(d-2)}\left(g^{\sst\L}_{\m\l}\,\d_{\n}^{\r}-g^{\sst\L}_{\n\l}\,\d_{\m}^{\r}\right),
\ee
where $\L$ is the cosmological constant and $g^{\L}$ is the (A)dS metric.

By requiring the Lagrangian to be invariant under the transformation $\d^{\sst\L}_\e$, 
we determine the Einstein tensor as
\be
	\left(\cG^{\sst \L}_\syd\,\phi\right)_{\m\n,}{}^{\r}=-\frac12\,
	\d_{\m\n\a\b}^{\r\l\g\d}\,\nabla^{\a}\,\nabla_{\l}\,\phi_{\g\d,}^{\quad\b}\,,
	\label{Einstein sym}
\ee
which has exactly the same form as the flat space one \eqref{Kr flat} except for
 the replacement of $\partial_\m$ by $\nabla_\m$\,. Note that, as opposed to flat space case, the order of derivatives is important in the expression \eqref{Einstein sym}.
This system admits a gauge-for-gauge symmetry
\be 
	\e_{\m\n}=\left(\nabla_{(\m}\,\nabla_{\n)}+\frac{2\L}{(d-1)(d-2)}\,g^{\sst \L}_{\m\n}\right)\zeta\,,
\ee 
which corresponds to the partially-massless spin-two transformation.
 
On the other hand, for the invariance under the transformation $\d^{\sst\L}_\th$\,,
the Einstein operator should acquire a mass-like term,
\be
	\cG^{\sst \L}_{\ayd}=
	\cG^{\sst \L}_{\syd}-\,m_\L^2
	\,\cI,\label{Einstein antisym}
\ee
where the mass-squared parameter is fixed by
\be
	m_\L^2=-\frac{4(d-3)}{(d-1)(d-2)}\,\L\,,
	\label{mass term}
\ee
and the mass-term operator by
\be
	(\cI\,\phi)_{\m\n,}{}^{\r}=\frac12\,\d_{\m\n\a}^{\r\b\g}\,\phi_{\b\g,}{}^{\a}
=\phi_{\m\n,}^{\quad\r}-2\,\d^{\,\r}_{[\n}\,\phi_{\m]\a,}{}^{\a}\,.
\ee
Both of the actions $\cG^{\sst \L}_{\syd}$ and $\cG^{\sst \L}_{\ayd}$ vanish identically for $d\leq 3$. 

Notice that the mass-squared parameter is positive in AdS and negative in dS.
In both cases, the equation with higher mass-squared term is unitary
--- $\cG^{\sst \L}_{\ayd}$ for AdS and $\cG^{\sst\L}_{\syd}$ for dS ---
whereas the other is non-unitary. See \cite{Zinoviev:2002ye} for more details.

To recapitulate, the simple hook field in (A)dS space cannot have a two-derivative
action respecting both of the gauge symmetries \eqref{hk sym},
but either the action $\cS^{\sst \L}_{\sst\rm E\,\syd}$ with only the symmetric parameter gauge symmetry
or the action $\cS^{\sst \L}_{\sst\rm E\,\ayd}$ with only the anti-symmetric parameter gauge symmetry:
\be 
	\cS^{\sst \L}_{\sst\rm E\,\syd}[\phi]=\la\phi\,|\,\cG^{\sst \L}_{\syd}\,\phi\ra\,,
	\qquad
	\cS^{\sst \L}_{\sst\rm E\,\ayd}[\phi]=\la\phi\,|\,\cG^{\sst \L}_{\ayd}\,\phi\ra\,.
	\label{Einstein hk}
\ee
Let us  examine the DoF of the above systems. We first
 consider the action  $\cS^{\sst \L}_{\sst\rm E\,\syd}$\,.
 The DoF of the system can be counted
 in terms of the $GL(d)$ Lorentz covariant tensors as
\ba
  \cS^{\sst \L}_{\sst\rm E\,\syd}\quad : \quad && \yng(2,1)_{\sst GL(d)} \ominus
   2\,\yng(2)_{\sst GL(d)}\oplus 3\,\bullet \nn
   &&\hspace{50pt}
   \ominus\   \yng(1,1)_{\sst GL(d)}
   \oplus\yng(1)_{\sst GL(d)}\,,
 \ea
 which has a non-trivial pattern due to the mixture of first- and second-class constraints
 as well as gauge-for-gauge.
Instead, in the Hamiltonian analysis, we find a
simpler pattern,
 \be
	 \yng(2,1)_{\sst SO(d-1)} \ominus \yng(2)_{\sst SO(d-1)}\oplus \bullet\,,
\ee
where we simply remove once the DoF of the traceless gauge parameter
and put back that of gauge-for-gauge.
We can further decompose these DoF as propagation on light-cone, ending up with
\be
	\cS^{\sst \L}_{\sst\rm E\,\syd}\quad:\quad \yng(2,1)_{\sst SO(d-2)}\oplus 
	\yng(1,1)_{\sst SO(d-2)}\,.
\ee
Similarly,  the number of DoF of the action $\cS^{\sst \L}_{\sst\rm E\,\ayd}$ 
can be counted in terms of $GL(d)$ covariant tensors as
 \ba
  \cS^{\sst \L}_{\sst\rm E\,\ayd}\quad : \quad  && \yng(2,1)_{\sst GL(d)} \ominus 
 2\,\yng(1,1)_{\sst GL(d)}
  \nn 
  &&\hspace{50pt} \ominus\ 
   \yng(2)_{\sst GL(d)}\oplus \yng(1)_{\sst GL(d)}\,.
  \ea
The Hamiltonian analysis gives the same result, in terms of $SO(d-1)$ branching:
\be
	\yng(2,1)_{\sst SO(d-1)} \ominus \yng(1,1)_{\sst SO(d-1)}\,.
\ee
In terms of light-cone DoF, the equation \eqref{Einstein antisym} propagates
\be
	\cS^{\sst \L}_{\sst\rm E\,\ayd}\quad:\quad \yng(2,1)_{\sst SO(d-2)}\oplus \yng(2)_{\sst SO(d-2)}\,.
\ee
Hence, compared to the flat space case, the hook action 
$\cS^{\sst \L}_{\sst\rm E\,\syd}$ describes extra DoF corresponding to the light-cone propagation of a
massless two-form field $\tiny\yng(1,1)$\,,
while  $\cS^{\sst \L}_{\sst\rm E\,\ayd}$ describes extra DoF of massless spin-two field $\tiny\yng(2)$\,.
The general pattern of the decomposition of modes in flat limit for
an arbitrary mixed-symmetry field has been conjectured in \cite{Brink:2000ag}
and proved in \cite{Boulanger:2008up,Boulanger:2008kw,Alkalaev:2009vm,Alkalaev:2011zv}.
Let us note that in four dimensions, the hook mode identically vanishes hence
the action $\cS^{\sst\L}_{\sst\rm E\,\ayd}$ describes only the massless spin two mode.
We will comment on this action in the section \ref{sec:3.1}, where a generalization to any dimensions with an off-shell field of type $[d-2,1]$ will be discussed in detail and the relation with the recent work \citep{Basile:2015jjd} will be clarified.

\subsection{Weyl Action}
\label{sec:2.3}

Even though there is no two-derivative action preserving both of the gauge symmetries of \eqref{hk sym} in (A)dS, there may exist a higher-derivative action which is invariant with respect to both of the symmetries.

It turns out that the four-derivative action 
invariant under both symmetries has a simple form,
\be
	\cS^{\sst \L}_{\rm\sst W}[\phi]
	=-\la \phi\,|\,\cG^{\sst\L}_{\syd}\,\cI^{-1}\,\cG^{\sst\L}_{\ayd}\,\phi\ra\,,
	\label{S4}
\ee
where the overall sign is chosen for the positive definite Euclidean action.
This action makes use of both Einstein tensors
as well as the inverse mass-term operator  $\cI^{-1}$ given by
\be
	(\cI^{-1}\,\phi)_{\m\n,}{}^{\r}
	=\phi_{\m\n,}{}^{\r}-\frac{2}{d-2}\,\d^{\,\r}_{[\n}\,\phi_{\m]\a,}{}^{\a}\,.
\ee
All the operators  $\cG^{\sst\L}_{\syd}$, $\cG^{\sst\L}_{\ayd}$ and $\cI$ are self-adjoint with respect to 
the scalar product \eqref{scalar pr}\,:
\be
	\la f\,|\,\cO\,g\ra=\la \cO\,f\,|\,g\ra\,,
	\qquad 
	\cO=\cG^{\sst \L}_{\syd},\ \cG^{\sst \L}_{\ayd},\  \cI\,.
\ee
Since the Einstein operators $\cG^{\sst\L}_\syd$ and $\cG^{\sst\L}_\ayd$ differ by 
a mass-like term,
it is easy to show that
\ba
	\cS^{\sst \L}_{\sst\rm W}[\phi]
	\eq - \la\phi\,|\left(\cG^{\sst \L}_{\syd}\,\cI^{-1}\,\cG^{\sst \L}_{\syd}
	-m_\L^2\,\cG^{\sst \L}_{\syd}\right)\phi\ra \nn
	\eq - \la\phi\,|\left(\cG^{\sst \L}_{\ayd}\,\cI^{-1}\,\cG^{\sst \L}_{\ayd}
	+m_\L^2\,\cG^{\sst \L}_{\ayd}\right)\phi\ra\,.
	\label{S4m}
\ea
In the first line, the action is manifestly invariant under the gauge symmetry $\d^{\sst\L}_\e$
with the symmetric parameter,
whereas the second line is manifestly invariant under the $\d^{\sst\L}_\th$ one
with anti-symmetric parameter.

In the flat limit, the  four-derivative action \eqref{S4} reduces to
\be 
	\cS^{\rm\sst flat}_{\rm\sst W}[\phi]=-\la\phi\,|\,\cG\,\cI^{-1}\,
	\cG\,\phi\ra\,,
	\label{S4 flat}
\ee
which is actually the unique action invariant under both gauge symmetries 
as well as the Weyl transformation,
\be
	\delta_\a\,\phi_{\m\n,\r}=\eta_{\r[\m}\,\a_{\n]}\,.
\ee
This is the analogue of the Weyl gravity action, which is uniquely fixed by diffeomorphism and Weyl symmetries.
Hence, one can regard $\cS^{\sst\rm flat}_{\sst\rm W}$ as 
the Weyl action for the simple hook field.
In four dimensions, this action has the conformal invariance,
and its form has been determined in \cite{Metsaev:2015rda}
together with other two-column mixed symmetry fields.

Coming back to the AdS action $\cS^{\sst\L}_{\sst\rm W}$ of \eqref{S4},
 one may wonder whether it
also admits a Weyl symmetry:
\be 
	\delta^{\sst\L}_{\a}\,\phi_{\m\n,\r}=g^{\sst\L}_{\r[\m}\,\a_{\n]}\,.
	\label{Weyl}
\ee
One can check the invariance of \eqref{S4} under \eqref{Weyl}
by a brute force computation,
but there is in fact a simpler way to see it. Since any linear combination of gauge symmetries should be a gauge symmetry, taking the following gauge parameters,
\ba
\e_{\m\n}(\a)=\nabla_{(\m}\,\a_{\n)}\,,\qquad
\th_{\m\n}(\a)=-\frac13\,\nabla_{[\m}\,\a_{\n]}\,,
\ea
we immediately get the Weyl transformation,
\ba
(\d_{\e(\a)}+\d_{\th(\a)})\,\phi_{\m\n,\r}
=\frac{\L}{(d-1)(d-2)}\,g^{\sst\L}_{\r[\m}\,\a_{\n]}\,.
\ea
This is due to the cancellation of derivative terms.
Hence,  in a sense,
the Weyl symmetry arises
as the (A)dS remnant of the gauge-for-gauge symmetry
in flat space \eqref{gfg}.
We conclude that {\it any theory of hook field in (A)dS, which is invariant with respect to both gauge symmetries  is also Weyl invariant}.

This conclusion can be generalized to any mixed symmetry fields in (A)dS. 
Therefore, any theory, that is invariant with respect to two distinct gauge transformations of mixed-symmetry field, has to enjoy certain Weyl symmetry.

This argument is true even for symmetric fields. 
Let us take the example of spin-two field $\phi_{\m\n}$\,: 
if there exists a theory of $\phi_{\m\n}$
invariant with respect to both massless and partially-massless
gauge symmetries,
\be
\d \phi_{\m\n}=\nabla_{(\m}\e_{\n)}+\nabla_{\m}\nabla_{\n}\s+\frac{2\L}{(d-1)(d-2)}\,
g^{\L}_{\m\n}\,\s\,,
\ee
then such a theory will also admit a Weyl symmetry.
For the demonstration, 
it is enough to set the gauge parameter $\e_\m=-\nabla_\m\s$
to get  $\d \phi_{\m\n}=\frac{2\L}{(d-1)(d-2)}\,g^{\L}_{\m\n}\,\s$. The action having both of massless and partially-massless gauge symmetries can be constructed analogously to \eqref{S4}, 
using the massless Einstein operator $\cG^{\sst \L}_m$
and the partially-massless one $\cG^{\sst \L}_{pm}$\,, as
\be
	\cS_{\rm\sst W}^{\sst \L}[\phi]
	=-\la \phi\,|\,\cG^{\sst \L}_{\sst m}\,
	\cI_{\rm\sst FP}^{-1}\,\cG^{\sst \L}_{\sst pm}\,\phi\ra\,,
	\label{conformal gravity}
\ee
where $\cG^{\sst \L}_{pm}$ differs $\cG^{\sst \L}_m$ by
a particular mass term given through a Fierz-Pauli operator $\cI_{\rm\sst FP}$\,.
This four-derivative action \eqref{conformal gravity}
is nothing but Conformal Gravity linearized around (A)dS background. 
As the reader might notice,
this is the inverse logic
of the discussion in \cite{Deser:2012qg}. In general, one can use this argument to support the conjecture of \cite{Joung:2012qy} about the spectrum of higher-derivative Weyl-like actions for symmetric higher spin fields.

\subsubsection{Curvature Formulation}

In order to grasp some geometrical intuitions, 
let us revisit our construction in terms of (generalized) curvatures.
First of all, notice that an analogue of the Riemann curvature can be constructed for the hook field.
It is a tensor having the symmetry of the $GL(d)$ Young diagram $\tiny\yng(2,2,1)$  given in flat space by
\be
(\cR\,\phi)_{\m_1\m_2\m_3,}{}^{\n_1\n_2}=6\,\partial^{[\n_1}\,\partial_{[\m_1}\,\phi_{\m_2\m_3],}{}^{\n_2]}\,,
\ee
and is invariant with respect to both gauge symmetries
\eqref{flat gs}. Around (A)dS background,
it is impossible to construct a curvature
invariant under both gauge transformations \eqref{hk sym}.
Instead, we can consider the curvature,
\be
(\cR^{\sst\L}_{\syd}\,\phi)_{\m_1\m_2\m_3,}{}^{\n_1\n_2}
=6\,\nabla^{[\n_1}\,\nabla_{[\m_1}\,\phi_{\m_2\m_3],}{}^{\n_2]}\,,
\label{Curv sym}
\ee
which is invariant under the gauge transformation
$\delta^{\sst \L}_{\e}$ with symmetric parameter.
In order to get a curvature that is invariant 
with respect to the gauge symmetry $\delta^{\sst \L}_{\th}$ with anti-symmetric parameter,
one should deform the above curvature into 
\be
(\cR^{\sst\L}_{\ayd}\,\phi)_{\m_1\m_2\m_3,}{}^{\n_1\n_2}
=(\cR^{\sst\L}_{\syd}\,\phi)_{\m_1\m_2\m_3,}{}^{\n_1\n_2}
+\frac{24\,\L}{(d-1)(d-2)}\,\d_{[\m_1}^{[\n_1}\,\phi_{\m_2\m_3],}{}^{\n_2]}\,.\label{Curv antisym}
\ee
These two curvatures can be directly related to the Einstein actions \eqref{Einstein hk}:
the Einstein tensors \eqref{Einstein sym} and \eqref{Einstein antisym} are 
 given through the curvatures
\eqref{Curv sym} and \eqref{Curv antisym} respectively as 
\be
(\cG^{\sst\L}_{\syd}\,\phi)_{\m\n,}{}^\r
=-\frac1{12}\,\d_{\m\n\a\b}^{\r\g\d\s}\,(\cR^{\sst \L}_{\syd}\,\phi)_{\g\d\s,}{}^{\a\b}\,,
\qquad
(\cG^{\sst\L}_{\ayd}\,\phi)_{\m\n,}{}^\r
=-\frac1{12}\,\d_{\m\n\a\b}^{\r\g\d\s}\,(\cR^{\sst \L}_{\ayd}\,\phi)_{\g\d\s,}{}^{\a\b}\,,
\ee
in the same manner as Einstein tensor is given through the Riemann curvature: remind, that Einstein tensor can be written as $G_{\m}^{\n}=R_{\m}^{\n}-\tfrac12\, \d_{\m}^{\n}\,R
=-\frac14\, \d_{\m\l\r}^{\n\a\b}\,R_{\a\b,}^{\quad\,\l\r}$, where $R_{\a\b,\l\r}$ is Riemann curvature, $R_{\m\n}$ and $R$ are Ricci curvature and its trace respectively.
Hence, it is clear how each curvature is related to the two-derivative Einstein
action for the simple hook field.

Let us now move on  to the four-derivative action $\cS^{\sst\L}_{\sst\rm W}$\,,
which admits Weyl symmetry. In the case of gravity, the Weyl gravity action
is simply given by the square of  Weyl tensor.
Hence, the action $\cS^{\sst\L}_{\sst\rm W}$ may also admit such a description.
We can first define the Weyl tensor from the curvature
as the traceless part of the latter. This can be conveniently done
by introducing the traceless projector $\cT$ whose explicit form
is not necessary for now.
In terms of $\cT$,  the Weyl tensor can be determined as
\be
	\cW=\mathcal T\,\cR_{\syd}^{\sst\L}=\mathcal T \,\cR_{\ayd}^{\sst\L}\,.
\label{WeylTensorForHook}
\ee
Remark that  the definition of the Weyl tensor 
does not distinguish between curvatures $R^{\sst\L}_\syd$ and $R^{\sst\L}_\ayd$
because the difference between two curvatures is precisely a trace term 
which is projected out under the action of $\mathcal T$.
 Therefore, the Weyl tensor is invariant under both gauge symmetries.
 As the invariance with respect to both gauge symmetries implies the invariance
 under Weyl symmetry transformation, the tensor $(\cW\,\phi)_{\m\n\r,\l\k}$
 can be rightfully referred to as Weyl tensor.
Again, mimicking the Weyl gravity case, 
we can consider a four-derivative action, that is the square of the Weyl tensor.
Since the four-derivative action invariant under both gauge symmetries is unique, 
they should be simply related with a proportionality constant as
\be
	\cS^{\sst \L}_{\sst\rm W}[\phi]
	=-\frac{d-3}{d-4}\,\la \cW\,\phi\,|\,\cW\,\phi\ra\,.
	\label{WeylSquaredHook}
\ee
The uniqueness of the four derivative action  with both symmetries in (A)dS can be checked directly, but can be also understood by the following simple argument: As we have shown above, any action for hook field with at most four derivatives and invariant with respect to both gauge symmetries in (A)dS enjoys Weyl symmetry, and therefore coincides with the action \eqref{S4 flat} in the flat limit. It follows then, that if there are distinct actions with those symmetries, their difference is encoded in two-derivative terms proportional to (A)dS curvature. Therefore there is a linear combination of them that is a two-derivative action invariant under both symmetries.
Since there is no such two-derivative action, it follows that 
the four-derivative action \eqref{S4} is unique.

In order to show that the formula \eqref{WeylSquaredHook}
coincides with \eqref{S4}, one needs to perform several integrations by part ignoring the boundary terms.
In the case of Weyl gravity, this amounts to
\be
	W^{\m\n,\r\l}\, W_{\m\n,\r\l} = \cL_{\rm\sst GB}+\frac{4(d-3)}{d-2}\left(
	R^{\m\n}\,R_{\m\n}-\frac{d}{4(d-1)}\,R^2\right)\,,
\ee
where the Gauss-Bonet (GB) term given by
\be
	\cL_{\rm\sst GB}=\frac14\,\d_{\n_1}^{\m_1}{}_{\n_2}^{\m_2}{}_{\n_3}^{\m_3}{}_{\n_4}^{\m_4}\,
R_{\m_1\m_2,}{}^{\n_1\n_2}\,R^{\n_3\n_4,}{}_{\m_3\m_4}\,.
\ee
 vanishes identically in $d\le 3$ and is a topological invariant in $d=4$\,.
 In higher dimensions, its linearization  always reduces to a boundary term.
Here again, the `Weyl squared' action for the simple hook field
can be related to the form \eqref{S4}
by adding a GB-like term for the simple hook field.
In flat space, such term  is given by
\be
\mathcal{L}^{\rm\sst flat}_{\rm\sst GB}(\phi)
=\frac1{2!\,3!}\,\d_{\n_1}^{\m_1}{}_{\n_2}^{\m_2}{}_{\n_3}^{\m_3}{}_{\n_4}^{\m_4}{}_{\n_5}^{\m_5}\,
(\cR\,\phi)_{\m_1\m_2\m_3,}{}^{\n_1\n_2}\,(\cR\,\phi)^{\n_3\n_4\n_5,}{}_{\m_4\m_5}\,.
\label{GBHook}
\ee
This term now vanishes identically in $d\le 4$ and 
becomes a boundary term in $d\ge 5$\,.
There is a unique (A)dS generalization of \eqref{GBHook} that is invariant with respect to both gauge symmetries and 
reduces to a boundary term in $d\ge5$. It is given by
\be
\mathcal{L}^{\sst \L}_{\rm\sst GB}(\phi)
=\frac1{2!\,3!}\,\d_{\n_1}^{\m_1}{}_{\n_2}^{\m_2}{}_{\n_3}^{\m_3}{}_{\n_4}^{\m_4}{}_{\n_5}^{\m_5}\,
(\cR^{\sst\L}_\syd\,\phi)_{\m_1\m_2\m_3,}{}^{\n_1\n_2}\,(\cR^{\sst\L}_\ayd\,\phi)^{\n_3\n_4\n_5,}{}_{\m_4\m_5}\,.
\label{GBHook AdS}
\ee
The Weyl squared action \eqref{WeylSquaredHook}
differs from the action \eqref{S4} 
by the boundary term \eqref{GBHook AdS}:
\be
	\la \cW\,\phi\,|\,\cW\,\phi\ra=
	\frac{1}{12}\int d^dx\sqrt{|g^{\sst \L}|}\,
	\cL^{\sst\L}_{\rm\sst GB}(\phi)
	+\frac{d-4}{d-3}\,	\la \phi\,|\,\cG^{\sst\L}_{\syd}\,\cI^{-1}\,\cG^{\sst\L}_{\ayd}\,\phi\ra\,.\label{WeylGB}
\ee
Let us notice that both 
the Weyl tensor and the GB-like term \eqref{GBHook} vanish identically
for $d\leq 4$\,, while the action \eqref{S4}
does not vanish in $d=4$. The equation \eqref{WeylGB} still holds, since the coefficient in front of the last term vanishes for $d=4$\,. In fact, in four dimensions, the action \eqref{S4} describes special spectra, which is different from the off-shell field. We will come back to this later.

\subsubsection{Factorization and Degrees of Freedom}

The Lagrangian \eqref{S4} can be written in the ordinary-derivative form 
by introducing an auxiliary field $f_{\m\n,\r}$ as
\be
\cS^{\sst\L}_{\rm\sst W}[\phi]\simeq m_\L^2
\left(\la\phi\,|\, \cG^{\sst\L}_{\syd}\,\phi\ra
+2\,\la \phi\,|\,\cG^{\sst\L}_\syd\,f\ra
+m^2_\L\,\la f\,|\,\cI\,f\ra\right).\label{AuxWeyl}
\ee
We can solve the equation for $f$ and go back to the Lagrangian \eqref{S4}. 
The Lagrangian \eqref{AuxWeyl} can be diagonalized into
\be
\cS^{\sst\L}_{\sst\rm W}[\phi]\simeq m_\L^2\left(\cS^{\sst\L}_{\sst\rm E\,\syd}[h]
-\cS^{\sst\L}_{\sst\rm E\,\ayd}[f]\right).
\label{ODActionWeylAdS}
\ee
 using the  field redefinition,
\be\label{FieldRedef}
	\phi_{\m\n,\r}=h_{\m\n,\r}-f_{\m\n,\r}\,.
\ee
This rewriting procedure is singular in the flat limit (note that the action
\eqref{AuxWeyl} is not well defined in the flat space limit $m_\L^2\to 0$), while for the action \eqref{S4} the flat space limit is well defined and the number of DoFs of the theory is the same in (A)dS and flat spaces. 
The action \eqref{ODActionWeylAdS} contains both short representations 
of the hook in (A)dS$_d$.
In the flat-space limit, it propagates two relatively ghost hook-helicity $\tiny\yng(2,1)$ modes (which are propagating only for $d\ge5$), supplemented with a spin-two helicity mode $\tiny\yng(2)$ and a two-form helicity mode $\tiny\yng(1,1)$\,.
Even in (A)dS background, the reasoning in terms of 
the $SO(d-2)$ helicity modes  can serve as a useful guideline,
despite the fact that they are not irreducible anymore when $\L\neq 0$.

Let us now analyze the DoFs of the theory $\cS^{\sst \L}_{\rm\sst W}$ in terms
of the helicity modes.
The result of the analysis can be schematically formulated as
 \be
	m^2_{\sst \L} \left(\,\yng(2,1)_h + \,\L\,\yng(1,1)_h \right)
	-m^2_{\sst \L} \left(\,\yng(2,1)_f -\L\,\yng(2)_f\right).
	\label{DoF hf}
\ee
The $h$-field comes with a supplementary propagating two-form mode $\tiny\yng(1,1)_h$\,, 
which is unitary in dS space and 
non-unitary in AdS \cite{Zinoviev:2002ye}. Since the action for that mode comes with the factor $m_\L^2$ in front, which is positive in AdS and negative in dS, corresponding two-form $\tiny\yng(1,1)_h$ is non-unitary in both cases, while the leading hook $\tiny\yng(2,1)_h$ itself is unitary in AdS and non-unitary in dS. In the same way, the spin-two mode $\tiny\yng(2)_f$ of the $f$-field has the same sign of kinetic term as the leading hook $\tiny\yng(2,1)_f$ in AdS and opposite sign in dS, therefore is non-unitary in both cases, taking into account the factor $-m_\L^2$ in front of the action for $f$. We conclude, that among the four propagating helicity modes,  only 
one of the hook modes has positive sign of kinetic term, therefore the theory is non-unitary for any $d\geq 5$, when hook modes propagate.

In four dimensions, the helicity group is $SO(2)$, which does not have any hook representation, therefore the hook helicity modes do not propagate,
hence we are left with a pseudo scalar $\tiny\yng(1,1)_h\sim \bullet_h$ and a spin-two 
$\tiny\yng(2)_f$ modes both non-unitary.
We can render these two modes into unitary one by simply introducing a negative sign 
factor in the original action.
In this way, we can describe a unitary system of spin-two and scalar in four dimensions
 using the four-derivative Weyl action $\cS^{\sst \L}_{\sst\rm W}$\,.
This theory  is unitary in four dimensions and has the same spectrum 
for any value of $\L$  (including zero), and coincides with the massless limit of $4d$ New Massive Gravity \cite{Bergshoeff:2012ud} in flat space. 
This action can be equally obtained by a particular dimensional reduction
of massless spin-two field from $5d$ down to $4d$ \cite{Joung:2012sa}.

\subsubsection{Massive Deformation}

Let us consider mass deformations of the Weyl action 
 with Einstein terms.
 Again we have two options:
 the first one is to introduce $\cS^{\sst \L}_{\rm\sst E\,\syd}$ term
 so that the total action preserves the $\syd$ gauge symmetry:
\be 
	\cS^{\sst\rm massive}_{\syd}[\phi,m^2]
	= 	\cS^{\sst\L}_{\rm\sst W}[\phi]+m^2\,\cS^{\sst\L}_{\sst\rm E\,\syd}[\phi]\,,
\ee
and the other possibility is keeping the $\ayd$ symmetry by considering,
\be 
	\cS^{\sst\rm massive}_{\ayd}[\phi,m^2]
	= 	\cS^{\sst\L}_{\rm\sst W}[\phi]+m^2\,\cS^{\sst\L}_{\sst\rm E\,\ayd}[\phi]\,.
\ee
 In each cases, we can introduce an auxiliary field $f$ and do a proper redefinition
 of  the type `$\phi=h-f$' as in the Weyl action case
 so that  the four-derivative actions turn into two-derivative ones,
\ba
	&& \cS^{\rm\sst massive}_{\syd}[\phi,m^2]
	\simeq
	\left(m^2_\L+m^2\right)\left(\la h\,|\,\cG^{\sst \L}_{\syd}\,h\ra
	-\la f\,|\left(\cG^{\sst \L}_{\ayd}-m^2\,\cI\right) f\ra\right),\\
	&& \cS^{\rm\sst massive}_{\ayd}[\phi,m^2]\simeq 
	\left(m^2_\L-m^2\right)\left(\la h\,|\left(\cG^{\sst \L}_{\syd}-m^2\,\cI\right) h\ra
	-\la f\,|\,\cG^{\sst \L}_{\ayd}\,f\ra\right).
\ea
One can see that depending on which symmetry we decide to preserve,
the other field --- whose gauge symmetry is spoiled by the Einstein term --- acquires a mass.
Let us also note that  
the action $\cS^{\rm\sst massive}_{\syd/\ayd}$ becomes singular
when $\mp m^2$ approaches $m_\L^2$\,.
This corresponds to the coincidence limit of two spectra described by $h$ and $f$\,.

 Now let us focus on the four dimensional case. In the previous section,
 we have already remarked that the Weyl action $\cS^{\L}_{\sst\rm W}$ 
 describes a massless spin-two and a pseudo scalar in four dimensions.
 With the deformation of the Einstein term, the action 
 $\cS^{\rm massive}_{\syd}$ describes
 \be
 \cS^{\rm massive}_{\syd}\sim
 	\left(m^2_{\L}+m^2\right)
 	\left( \L\,\bullet_h -\, \yng(2)^{\sst SO(3)}_f\right),
 \ee
 whereas the $\cS^{\rm massive}_{\ayd}$ gives
  \be
 \cS^{\rm massive}_{\ayd}\sim
 	\left(m^2_{\L}-m^2\right)
 	\left(\,\yng(2)^{\sst SO(3)}_h +\L\, \yng(2)^{\sst SO(2)}_f\right),
 \ee
 where we have dualized all the modes except for the last one $\tiny\yng(2)^{\sst SO(2)}_f$
 and used the fact that the hook helicity mode identically vanishes in four dimensions.
Let us remark that 
in $ \cS^{\rm massive}_{\syd}$ the two modes ---
scalar $\bullet_h$ and massive spin two $\tiny\yng(2)^{\sst SO(3)}_f$ ---
have the same sign of the kinetic term only for $\L<0$\,, namely in AdS space. 
On the contrary,
the two modes of $ \cS^{\rm massive}_{\ayd}$ --- massive and massless spin two fields $\tiny\yng(2)^{\sst SO(3)}_h$ and 
$\tiny\yng(2)^{\sst SO(2)}_f$ ---
have the same sign in dS space ($\L>0$).
Hence, again by introducing an overall minus sign in these actions,
we obtain two models of four-derivative theories with unitary propagation.
It will be interesting to explore possible links of these four-derivative formulations
with existing models.
For instance, the field content of  $\cS^{\rm massive}_{\ayd}$
coincides with that of bigravity \cite{Hassan:2011zd} proposed by Hasan and Rosen.

\section{Generalizations}
\label{sec:3}

All the discussions about the simple hook field
can be straightforwardly generalized to more general 
fields having the symmetry of two-column Young diagrams.
The number of columns is playing the role of `spin' and 
the two-column fields show many common features 
with spin-two field.
In particular, their Weyl actions contain four derivatives. 
This considerably reduces 
the technical complexities
and allows us to perform the analysis explicitly.

The only qualitative difference of two-columns fields
with respect to the hook one rises in
the case where the two columns have the same height,
which we refer to as `long window' diagram.
Hence, we shall first consider
the non-window case in below
and then do a separate analysis for the long window case in the succeeding section.

\subsection{Two Columns}
\label{sec:3.1}


In this section, we will generalize the results of Section \ref{sec:2} for a tensor field
$\phi_{\m_1\cdots\m_p,}^{\qquad\quad\n_1\cdots\n_q}$
having the symmetry of  two column Young diagram:
\be
	\phi_{\m_1\cdots\m_p,}^{\qquad\quad\n_1\cdots\n_q} \sim\,
	\parbox{25pt}{\begin{tikzpicture}
	\draw (0,0) rectangle (0.4,2.4);
	\draw (0.4,2.4) -- (0.8,2.4) -- (0.8,0.8) -- (0.4,0.8);
	\node at (0.2,1.2){${\st p}$};
	\node at (0.6,1.6){${\st q}$};
	\end{tikzpicture}}\,,
	\label{pq field}
\ee
where we assume $p$ is strictly larger than $q$\,.

\subsubsection{Einstein Action}

The action for this field is \ed{schematically} given making use of the diagram, 
\be
	S_{\rm\sst E}[\phi]
	\sim \parbox{45pt}{\begin{tikzpicture}
	\draw (0,0) rectangle (0.4,2.4);
	\draw (0.4,2.4) -- (0.8,2.4) -- (0.8,0.8) -- (0.4,0.8);
	\draw (0.8,0.8) -- (0.8,-2) -- (0,-2) -- (0,0);
	\draw  (0.4,0) -- (0.4,-2);
	\draw (0,-0.4) -- (0.4,-0.4);
	\draw  (0.4,0.4) -- (0.8,0.4);
	\node at (0.2,1.2){${\st p}$};
	\node at (0.6,1.6){${\st q}$};
	\node at (0.2,-1.2){${\st q}$};
	\node at (0.6,-0.8){${\st p}$};
	\node at (0.2,-0.2){${\st \partial}$};
	\node at (0.6,0.6){${\st \partial}$};\,,
	\end{tikzpicture}}\,,
	\label{pq Lagrangian}
\ee
which is made of two fields and two derivatives
contracted with 
the generalized Kronecker delta $\d_{\m_1\cdots\m_{p+q+1}}^{\n_1\cdots\n_{p+q+1}}$\,. \ed{This form of the action for generic two column fields was introduced in \cite{Bekaert:2004dz}, generalizing the analogous formula for simple hook \cite{Curtright:1980yk} and the window diagram  \cite{Boulanger:2004rx}. It is equivalent to Labastida action \cite{Labastida:1986ft} up to total derivatives.}
The corresponding Einstein tensor is hence given by
\be
(\cG\,\phi)_{\m_1\cdots\m_p,}{}^{\n_1\cdots\n_q}
=-\frac{(-1)^{(p+1)(q+1)}}{p!\,q!}\,
\d_{\m_1\cdots\m_{p+q+1}}^{\n_1\cdots\n_{p+q+1}}\,
\partial^{\m_{p+1}}\,\partial_{\n_{q+1}}\,
\phi_{\n_{q+2}\cdots\n_{p+q+1},}{}^{\m_{p+2}\cdots\m_{p+q+1}}\,,
\label{pq einstein}
\ee
which is invariant with respect to two gauge symmetries:
\ba
&\d_\e\,\phi_{\m_1\cdots\m_p\,,}{}^{\n_1\cdots\n_q}
=\partial_{[\m_1}\,\e_{\m_2\cdots\m_p]\,,}{}^{\n_1\cdots\n_q}\,,\\
&\d_\th\,\phi_{\m_1\cdots\m_p\,,}{}^{\n_1\cdots\n_q}
=\partial^{[\n_1}\,\th_{\m_1\cdots\m_p\,,}{}^{\n_2\cdots\n_q]}
+(-1)^{p+1}\,\frac{p}{p-q+1}\,
\partial_{[\m_1}\,\th_{\m_2\cdots\m_p]}{}^{[\n_1\,,\,\n_2\cdots\n_q]},
	\label{pq gauge}
\ea
with gauge parameters of the following Young symmetry type:
\be
 \e_{\m_1\cdots\m_{p-1}\,,}{}^{\n_1\cdots\n_{q}}\sim\,
	\parbox{45pt}{\begin{tikzpicture}
	\draw (0,0) rectangle (0.4,2);
	\draw (0.4,2) -- (0.8,2) -- (0.8,0.4) -- (0.4,0.4);
	\draw [dotted] (0,0) rectangle (0.4,-0.4);
	\node at (-0.3,1){${\st p-1}$};
	\node at (0.6,1.2){${\st q}$};
	\end{tikzpicture}}\,,\qquad
 \th_{\m_1\cdots\m_{p}\,,}{}^{\n_1\cdots\n_{q-1}}\sim\,
	\parbox{40pt}{\begin{tikzpicture}
	\draw (0,0) rectangle (0.4,2.4);
	\draw (0.4,2.4) -- (0.8,2.4) -- (0.8,1.2) -- (0.4,1.2);
	\draw [dotted] (0.4,1.2) rectangle (0.8,0.8);
	\node at (0.2,1.2){${\st p}$};
	\node at (1.1,1.8){${\st q-1}$};
	\end{tikzpicture}}\,.
\ee
The Lagrangian \eqref{pq Lagrangian}, or 
equivalently the Einstein tensor
\eqref{pq einstein} 
can be also given through the gauge invariant curvature tensor $(\cR\,\phi)_{\m_1\cdots\m_{p+1}\,,}{}^{\n_1\cdots\n_{q+1}}$, which again 
has a symmetry of two-column Young diagram
but with additional boxes in each column (corresponding to derivatives):
\be
 (\cR\,\phi)_{\m_1\cdots\m_{p+1}\,,}{}^{\n_1\cdots\n_{q+1}}\sim\,
	\parbox{25pt}{\begin{tikzpicture}
	\draw (0,0) rectangle (0.4,2.4);
	\draw (0.4,2.4) -- (0.8,2.4) -- (0.8,0.8) -- (0.4,0.8);
	\draw (0,0) -- (0,-0.4) -- (0.4,-0.4) -- (0.4,0);
	\draw (0.4,0.4) -- (0.8,0.4) -- (0.8,0.8);
	\node at (0.2,1.2){${\st p}$};
	\node at (0.6,1.6){${\st q}$};
	\node at (0.2,-0.2){${\st \partial}$};
	\node at (0.6,0.6){${\st \partial}$};
	\end{tikzpicture}}\,.
	\label{curvature}
\ee
Its explicit expression reads
\be
	(\cR\,\phi)_{\m_1\cdots\m_{p+1},}{}^{\n_1\cdots\n_{q+1}}
=(p+1)(q+1)\,\partial_{[\m_1}\,\partial^{[\n_1}\,\phi_{\m_2\cdots\m_{p+1}],}
{}^{\n_2\cdots\n_{q+1}]}\,,
\label{pq curvature}
\ee
and it is invariant with respect to both gauge symmetries
\eqref{pq gauge}. 

Around (A)dS background, the curvature cannot be gauge invariant with respect to both of the symmetries, and one has to choose a zero-derivative deformation to preserve one of the symmetries:
\ba
&\d^{\sst\L}_\e\,\phi_{\m_1\cdots\m_p\,,}{}^{\n_1\cdots\n_q}
=\nabla_{[\m_1}\,\e_{\m_2\cdots\m_p]\,,}{}^{\n_1\cdots\n_q}\,,\\
&\d^{\sst \L}_\th\,\phi_{\m_1\cdots\m_p\,,}{}^{\n_1\cdots\n_q}
=\nabla^{[\n_1}\,\th_{\m_1\cdots\m_p\,,}{}^{\n_2\cdots\n_q]}
+(-1)^{p+1}\,\frac{p}{p-q+1}\,
\nabla_{[\m_1}\,\th_{\m_2\cdots\m_p]}{}^{[\n_1\,,\,\n_2\cdots\n_q]},
\label{pq gauge AdS}
\ea
The curvature invariant 
under the $\delta^{\sst \L}_{\e}$ symmetry 
takes the form,
\be
(\cR^{\sst \L}_{\sst [p-1,q]}\,\phi)_{\m_1\cdots\m_{p+1},}{}^{\n_1\cdots\n_{q+1}}
=(p+1)(q+1)\,\nabla^{[\n_1}\,\nabla_{[\m_1}\,
\phi_{\m_2\cdots\m_{p+1}],}{}^{\n_2\cdots\n_{q+1}]}\,,\label{CurvTheta}
\ee
which is again a simple extension
of the flat space curvature \eqref{pq curvature}.
Considering now the 
curvature invariant under the $\delta^{\sst \L}_{\th}$
transformation, we find
\ba
&& (\cR^{\sst\L}_{\sst [p,q-1]}\,\phi)_{\m_1\cdots\m_{p+1},}{}^{\n_1\cdots\n_{q+1}}\nn
&&=\,
(\cR^{\sst \L}_{\sst [p-1,q]}\,\phi)_{\m_1\cdots\m_{p+1},}{}^{\n_1\cdots\n_{q+1}}
-\frac{2(p-q+1)\,\L}{(d-1)(d-2)}\,\d_{[\m_1}^{[\n_1}\,\phi_{\m_2\cdots\m_{p+1}],}{}^{\n_2\cdots\n_{q+1}]}\nonumber\\
&&=\,(p+1)(q+1)\,\nabla_{[\m_1}\,\nabla^{[\n_1}\,\phi_{\m_2\cdots\m_{p+1}],}{}^{\n_2\cdots\n_{q+1}]}
-\frac{2\,\L}{(d-1)(d-2)}\,\d_{[\m_1}^{[\n_1}\,\phi_{\m_2\cdots\m_{p+1}],}{}^{\n_2\cdots\n_{q+1}]}\,.\qquad\quad\label{CurvSigma}
\ea
We can see that these two curvatures differ
by a zero-derivative term. 
One can then construct Einstein tensors out of these
curvatures as
\ba
	 (\cG^{\sst\L}_{\sst [p-1,q]}\,\phi)_{\m_1\cdots\m_{p},}{}^{\n_1\cdots\n_{q}}
	 \eq -\frac{(-1)^{(p+1)(q+1)}}{(p+1)!(q+1)!}\,\delta_{\m_1\cdots \m_{p+q+1}}^{\n_1\cdots \n_{p+q+1}}\,
	 (\cR^{\sst\L}_{\sst [p-1,q]}\,\phi)_{\n_{q+1}\cdots\n_{p+q+1},}{}^{\m_{p+1}\cdots\m_{p+q+1}}\,,\qquad\,\,\\
	  (\cG^{\sst\L}_{\sst [p,q-1]}\,\phi)_{\m_1\cdots\m_{p},}{}^{\n_1\cdots\n_{q}}
	 \eq -\frac{(-1)^{(p+1)(q+1)}}{(p+1)!(q+1)!}\,\delta_{\m_1\cdots \m_{p+q+1}}^{\n_1\cdots \n_{p+q+1}}\,
	 (\cR^{\sst\L}_{\sst [p,q-1]}\,\phi)_{\n_{q+1}\cdots\n_{p+q+1},}{}^{\m_{p+1}\cdots\m_{p+q+1}}\,.\qquad\,\,
\ea
The Einstein tensors
immediately define the Einstein actions
\ba
\cS^{\sst\L}_{\sst {\rm E}\,[p-1,q]}=\la \phi\,|\,\cG^{\sst\L}_{\sst [p-1,q]}\,\phi\ra\,,\quad
\cS^{\sst\L}_{\sst {\rm E}\,[p,q-1]}=\la \phi\,|\,\cG^{\sst\L}_{\sst [p,q-1]}\,\phi\ra\,
\label{El}
\ea
 for the  two-column field \eqref{pq field}.
The action $\cS^{\sst\L}_{\sst {\rm E}\,[p-1,q]}$
is (not) unitary around (A)dS
and the opposite for $\cS^{\sst\L}_{\sst {\rm E}\,[p,q-1]}$\,.
When \mt{p+q=d-1}, the $[p,q]$ mode of 
the Einstein actions  $\cS^{\sst\L}_{\sst {\rm E}\,[p-1,q]}$
or $\cS^{\sst\L}_{\sst {\rm E}\,[p,q-1]}$
vanishes identically leaving only the $[p,q-1]$ or $[p-1,q]$ mode, respectively.
The kinetic terms of these modes come with the factor $\L$ or $-\L$\,.
With an appropriate choice of an overall factor of the Einstein action, 
these modes may become unitary.

\subsubsection*{Long hook $[d-2,1]$}

An interesting example of the case $p+q=d-1$  is  the \emph{long hook} where $p=d-2$ and $q=1$\,.
The action $\rm{sign}(\L)\,\cS^{\sst\L}_{\sst {\rm E}\,[d-3,1]}$  has a propagating degree of freedom of a scalar,
whereas the other action $-\,\rm{sign}(\L)\,\cS^{\sst\L}_{\sst {\rm E}\,[d-2,0]}$  has a propagating 
mode dual to massless graviton.

The long hook field $[d-2,1]$ is used in New Massive Gravity in arbitrary dimensions \cite{Joung:2012sa,Dalmazi:2012dq},
and also in the recent work  \cite{Basile:2015jjd}.
Two different ways the long hook field appears in \cite{Basile:2015jjd} and   \cite{Joung:2012sa} 
have analogous features which deserve a few remarks.
On the one hand in  \cite{Basile:2015jjd}, the authors obtained a two-derivative action 
from the linearized Einstein-Cartan gravity in AdS by integrating out the vielbein instead 
of the spin connection. This determines the linearized vielbein in terms of spin connection as 
$h_\mu^a=h_\m^a(\o)$\,, which has the form of Schouten tensor written in terms of spin connection. 
On the other hand in \cite{Joung:2012sa},
the simple hook field $\phi_{\m\n,\r}$ appears as a result of solving the constraint,
\be
	\partial^\m\,\partial^\n\,h_{\m\n}-\Box\,h^\m{}_\m=0\,,
	\label{constraint h}
\ee
 arising in the course of a special dimensional reduction of the massless spin two system.
The solution to \eqref{constraint h} is given \cite{DuHe} in terms of a hook field $\varphi_{\m\n,\r}$\,:
$h_{\m\n}=h_{\m\n}(\varphi)$\,.
Interestingly, the form $h_\m^a(\o)$ can be brought to the form of $h_{\m\n}(\varphi)$
using gauge transformations and with $\partial_\mu$ replaced by the AdS covariant derivative.
Moreover, it has been shown  \cite{Joung:2012sa} that the  action resulting after solving the constraint \eqref{constraint h} has additional symmetry $\d_\s\,\phi_{\m\n,\r}=\eta_{\r[\mu}\,\pr_{\nu]}\s$
under which $h_{\m\n}(\varphi)$ transforms like the Weyl transformation of Schouten tensor,
 $\d_\s\,h_{\m\n}(\varphi)=\na_\m\pr_\nu\s$. 
 This interplay between two constructions in  \cite{Basile:2015jjd} and \cite{Joung:2012sa}  can be understood from 
 the fact that the constraint \eqref{constraint h} is the identity 
 that Schouten tensor satisfies.
Let us notice also that  the action of  \cite{Basile:2015jjd}
after dualization actually coincides with $\cS^{\sst\L}_{\sst {\rm E}\,[d-2,0]}$\,\ed{, the second action of \eqref{El} for $p=d-2\,,\, q=1$}.
Another interesting feature of the construction \cite{Joung:2012sa} is that 
the final action is given by massive Fierz-Pauli action in terms of $h_{\m\n}(\varphi)$\,.
After dualization, in terms of the dual long hook field, $\phi_{\mu_1\cdots\mu_{d-2},\nu}$,
the two-derivative part proportional to $h^{[\m}_{\m}(\varphi)\,h^{\n]}_{\n}(\varphi)$
 coincides with the Einstein action $\cS_{\sst {\rm E}}[\phi]$,
while the four-derivative one, proportional to $\la h(\varphi)\,|\,\cG\,h(\varphi)\ra$, 
coincides with the Weyl action $\cS_{\sst\rm W}[\phi]$.

\subsubsection{Weyl Action}

Let us now move to the Weyl action.
For that, we consider
the Weyl tensor for two column field \eqref{pq field}
by making use of the trace projector $\cT$ as
\be
	\cW=\mathcal T\,\cR_{\sst [p-1,q]}^{\sst\L}=\mathcal T \,\cR_{\sst [p,q-1]}^{\sst\L}\,.
\label{WeylTensorForpq}
\ee
The form of the trace projector
acting on a $[p+1,q+1]$ tensor can be conveniently given by the following expression:
\be
\mathcal T = \frac{(-1)^{(p+1)(q+1)}}{(p+1)!(q+1)!}\left(\d_{p+q+2}-\frac{d-p-q-1}{d-p-q}\d_{p+q+1}(\d_{p+q})^{-1}\d_{p+q+1}\right),
\label{TraceProjector pq}
\ee
where the operator $\d_n$ acts on the field $f_{\m_1\cdots\m_{p}\,,}{}^{\n_1\cdots\n_{q}}$ of type $[p,q]$ in the following way
\be
(\d_n f)_{\r_{1}\cdots\r_{n-q}\,,}{}^{\l_{1}\cdots\l_{n-p}}
=\d^{\l_1\cdots\l_{n-p}\m_1\cdots\mu_p}_{\r_1\cdots\r_{n-q}\n_1\cdots\n_{q}}\,
f_{\m_{1}\cdots\m_{p}\,,}{}^{\n_{1}\cdots\n_{q}}\,,
\ee
contracting all of its indices to the generalized Kronecker-delta. Inverse operator $(\d_{p+q})^{-1}$ is defined as:
\be
(\d_{p+q}\,(\d_{p+q})^{-1}\,f)_{\m_1\cdots\m_{p}\,,}{}^{\n_1\cdots\n_{q}}=
f_{\m_1\cdots\m_{p}\,,}{}^{\n_1\cdots\n_{q}}=
((\d_{p+q})^{-1}\,\d_{p+q}\,f)_{\m_1\cdots\m_{p}\,,}{}^{\n_1\cdots\n_{q}}\,.
\ee
The tracelessness of the expression 
\eqref{TraceProjector pq} can be proven using the identity,
\be
\d_{\n_1}^{\m_1}\,\d_{\m_1\dots\m_{n}}^{\n_1\dots\n_{n}}
=(d-n+1)\,\d_{\m_2\dots\m_{n}}^{\n_2\dots\n_{n}}\,.
\ee
It is again clear that Weyl tensor \eqref{WeylTensorForpq} is 
invariant under both gauge symmetries \eqref{pq gauge AdS}
hence also under Weyl transformation:
\be
	\delta_\a\,\phi_{\mu_1\cdots\mu_p,}{}^{\n_1\cdots\n_q}
	=\delta_{[\mu_1}^{[\nu_1}\,
	\a_{\mu_2\cdots \mu_p],}{}^{\nu_2\cdots \nu_q]}\,.
	\label{Weyl sym}
\ee
Analogously to the hook case, 
the Weyl action is given by the  square
of Weyl tensors,
\be
\cS^{\sst \L}_{\rm\sst W}[\phi]= -\frac{d-p-q}{d-p-q-1}
\la \cW\,\phi\,|\,\cW\,\phi\ra\,,
\label{Sw W}
\ee
where the coefficient
is fixed such that
the Weyl action takes another representation,
\be
\cS^{\sst \L}_{\rm\sst W}[\phi]=-
\la \phi\,|\,\cG^{\sst\L}_{\sst [p-1,q]}\,\cI^{-1}\,
\cG^{\sst\L}_{\sst [p,q-1]}\,\phi\ra\,,
\label{Sw E}
\ee
up to a GB-like total derivative term,
\be
\cL^{\sst \L}_{\rm\sst GB}(\phi)
= (-1)^{(p+1)(q+1)}\,
\la\, \cR^{\sst\L}_{\sst [p-1,q]}\,\phi\,|\,
\delta_{p+q+2}\,\cR^{\sst\L}_{\sst [p,q-1]}\,\phi\,\ra\,.
\ee
In \eqref{Sw E} the mass operator $\cI$ is defined as
\be
\cI=\frac{(-1)^{p\,q}}{p!\,q!}\,\d_{p+q}\,.
\ee
Hence, one can see that the entire
construction of the hook example
can be generalized to the two column case
in a straightforward manner.

The DoF can be conveniently analyzed by using the 
same factorization technique as the simple hook case.
Skipping the straightforward derivation part, let us 
directly present the end  result,
 \be
	m^2_{\sst \L} \left(\,
	\parbox{22pt}{\begin{tikzpicture}
	\draw (0,0) rectangle (0.4,2.4);
	\draw (0.4,2.4) -- (0.8,2.4) -- (0.8,0.8) -- (0.4,0.8);
	\end{tikzpicture}}{\phantom{\Bigg|}}^{\sst SO(d-2)}_h
	 + \,\L\ 
	\parbox{22pt}{\begin{tikzpicture}
	\draw (0,0) rectangle (0.4,2.4);
	\draw (0.4,2.4) -- (0.8,2.4) -- (0.8,1.2) -- (0.4,1.2);
	\draw [dotted] (0.4,1.2) rectangle (0.8,0.8);
	\end{tikzpicture}}{\phantom{\Bigg|}}^{\sst SO(d-2)}_h \right)
	-m^2_{\sst \L} \left(\,
	\parbox{22pt}{\begin{tikzpicture}
	\draw (0,0) rectangle (0.4,2.4);
	\draw (0.4,2.4) -- (0.8,2.4) -- (0.8,0.8) -- (0.4,0.8);
	\end{tikzpicture}}{\phantom{\Bigg|}}^{\sst SO(d-2)}_f -\L\ 
	\parbox{22pt}{\begin{tikzpicture}
	\draw (0,0) rectangle (0.4,2);
	\draw (0.4,2) -- (0.8,2) -- (0.8,0.4) -- (0.4,0.4);
	\draw [dotted] (0,0) rectangle (0.4,-0.4);
	\end{tikzpicture}}{\phantom{\Bigg|}}^{\sst SO(d-2)}_f\right),
\ee
which is the natural generalization of the result \eqref{DoF hf}.
All these DoF vanish in dimensions $d\le p+q$\,.
When $d=p+q+1$\,, the helicity mode $[p,q]$ vanishes,
but there remains propagating DoFs given by
 \be
	-\parbox{42pt}{\begin{tikzpicture}
	\draw (0,0) rectangle (0.8,1.2);
	\draw (0.4,0) -- (0.4,1.2);
	\draw [dotted] (0,0) rectangle (0.4,-1.2);
	\draw [dotted] (0.4,0) rectangle (0.8,-0.4);
	\node at (-0.3,0.6){${\st q-1}$};
	\end{tikzpicture}}{\phantom{\Bigg|}}^{\sst SO(d-2)}_h 
	-\, 
	\parbox{22pt}{\begin{tikzpicture}
	\draw (0,0) rectangle (0.8,1.6);
	\draw (0.4,0) -- (0.4,1.6);
	\draw [dotted] (0,0) rectangle (0.4,-0.8);
	\node at (0.2,0.8){${\st q}$};
	\node at (0.6,0.8){${\st q}$};
	\end{tikzpicture}}{\phantom{\Bigg|}}^{\sst SO(d-2)}_f\,,
\ee
where we have dualized both modes.
Note that these modes have the same kinetic term sign,
hence can describe unitary propagation by introducing a negative factor
in the original action.
Notice that when $q=1$ we get in this way 
the DoF of
a scalar and a helicity two mode.
The corresponding $[p,q]=[d-2,1]$ Weyl action coincides 
in fact with the massless limit of New Massive Gravity action in any $d$
\cite{Joung:2012sa}, discussed in previous subsection.

\subsection{Long Window}
\label{sec:3.2}

In the special case where the height of two columns are equal, 
that is $q=p$\,:
\be
	\phi_{\m_1\cdots\m_p\,,}^{\qquad\quad\n_1\cdots\n_p} \sim\,
	\parbox{25pt}{\begin{tikzpicture}
	\draw (0,0) rectangle (0.8,2);
	\draw  (0.4,0) -- (0.4,2);
	\node at (0.2,1){${\st p}$};
	\node at (0.6,1){${\st p}$};
	\end{tikzpicture}}\,,
	\label{pp field}
\ee
the analysis is no more analogous to the hook field case,
but actually more similar to the spin-two case.
Let us see explicitly how this works.
First of all, the field $\phi_{\mu_1\cdots \m_p,}{}^{\n_1\cdots \n_p}$ admits only the gauge symmetry generated by 
the parameter of $[p,p-1]$ Young diagram,
because there is no Young diagram $[p-1,p]$.
Therefore, even in flat space, 
the long window has only one gauge symmetry.

In (A)dS background,
we can first consider the curvature  
$\cR^{\sst\L}_{\sst [p,p-1]}$ \eqref{CurvSigma}
or equivalently
the Einstein tensor $\cG^{\sst\L}_{\sst [p,p-1]}$
having the $[p,p-1]$ gauge symmetry.
In this case, there is nothing different from the generic two column case
and we can obtain the corresponding tensor and action.

Considering now the curvature 
$\cR^{\sst\L}_{\sst [p-1,p]}$ \eqref{CurvTheta}
or the Einstein tensor $\cG^{\sst\L}_{\sst [p-1,p]}$,
we first note that they cannot have the $[p-1,p]$ gauge symmetry as it simply does not exist.
Instead, one may wonder whether this action still plays a special role.
It tuns out that with the sacrifice of  the $[p-1,p]$ gauge symmetry,
the action  $\cS^{\sst \L}_{\sst {\rm E}\, [p-1,p]}$ acquires a new gauge symmetry,
\be
\d^{\sst \L}_\e\,\phi_{\m_1\cdots\m_p\,,}{}^{\n_1\cdots\n_p}
=\left(\nabla_{[\m_1}\,\nabla^{[\n_1}-
\frac{2\,\L}{(d-1)(d-2)}\,\d_{[\m_1}^{[\n_1}\right)
\e_{\m_1\cdots\m_{p-1}\,,}{}^{\n_1\cdots\n_{p-1}}\,,
\label{pp pm sym}
\ee
with the gauge parameter,
\be
	\e_{\m_1\cdots\m_{p-1}\,,}^{\qquad\quad\n_1\cdots\n_{p-1}} \sim\,
	\parbox{65pt}{\begin{tikzpicture}
	\draw (0,0) rectangle (0.8,1.6);
	\draw  (0.4,0) -- (0.4,1.6);
	\draw [dotted] (0,0) rectangle (0.4,-0.4);
	\draw [dotted] (0.8,0) -- (0.8,-0.4) --  (0.4,-0.4);
	\node at (-0.4,0.8){${\st p-1}$};
	\node at (1.2,0.8){${\st p-1}$};
	\end{tikzpicture}}\,.
	\label{pp pm gauge}
\ee
This is clearly the analogue of the partially-massless gauge symmetry of symmetric second rank field.
Indeed, when $p=1$ the corresponding action 
 coincides with that of partially-massless spin two.
Now considering the $[p-1,p-1]$ gauge symmetry \eqref{pp pm sym},
we can construct a one-derivative
gauge invariant, or curvature, as
\be
(\cC\,\phi)_{\m_1\cdots\m_{p+1}\,,}{}^{\n_1\cdots\n_p}
=(p+1)\,\nabla_{[\m_1}\,\phi_{\m_2\cdots\m_{p+1}],}{}^{\n_1\cdots\n_p},
\ee
and the action $\cS^{\sst \L}_{\sst {\rm E}\, [p-1,p]}$
can be also expressed as the square of this curvature and its traces.

We can proceed to construct a four-derivative action $\cS^{\sst\L}_{\sst\rm W}$
having
both of $[p,p-1]$ and $[p-1,p-1]$ gauge symmetries:
\ba
	\d^{\sst \L}\,\phi_{\m_1\cdots\m_p\,,}{}^{\n_1\cdots\n_p}
\eq \left(\nabla_{[\m_1}\,\nabla^{[\n_1}-\frac{2\,\L}{(d-1)(d-2)}\,\d_{[\m_1}^{[\n_1}\right)
\e_{\m_1\cdots\m_{p-1}\,,}{}^{\n_1\cdots\n_{p-1}}+\nn
&&+\,\nabla^{[\n_1}\,\th_{\m_1\cdots\m_p\,,}{}^{\n_2\cdots\n_p]}
+(-1)^{p+1}\,p\,
\nabla_{[\m_1}\,\th_{\m_2\cdots\m_p]}{}^{[\n_1\,,\,\n_2\cdots\n_p]}.
\label{pp gauge sym}
\ea
Such an action will be automatically invariant under the Weyl transformation \eqref{Weyl sym}
because it can be realized as a particular configuration of \eqref{pp gauge sym}.
The action $\cS^{\sst\L}_{\sst\rm W}$ can be constructed
exactly in the same way as in the generic $[p,q]$ case,
either using Weyl tensor as in \eqref{Sw W}
or using Einstein tensors as in \eqref{Sw E}.

The DoF of the Weyl action can be analysed in terms of $SO(d-2)$ representations in a similar manner as before.
The result reads
 \be
	m^2_{\sst \L} \left(\,
	\parbox{22pt}{\begin{tikzpicture}
		\draw (0,0) rectangle (0.8,2);
	\draw  (0.4,0) -- (0.4,2);
	\end{tikzpicture}}{\phantom{\Bigg|}}^{\sst SO(d-2)}_h
	 + \,\L\ 
	\parbox{22pt}{\begin{tikzpicture}
		\draw (0,0) rectangle (0.4,2);
	\draw  (0.4,0.4) -- (0.8,0.4) -- (0.8,2) -- (0.4,2);
	\draw [dotted] (0.4,0) rectangle (0.8,0.4);
	\end{tikzpicture}}{\phantom{\Bigg|}}^{\sst SO(d-2)}_h \right)
	-m^2_{\sst \L} \,
	\parbox{22pt}{\begin{tikzpicture}
	\draw (0,0) rectangle (0.8,2);
	\draw  (0.4,0) -- (0.4,2);
	\end{tikzpicture}}{\phantom{\Bigg|}}^{\sst SO(d-2)}_f \,,
\ee
Notice that the $h$- and $f$-fields
are   the long-window analogs of partially-massless spin two
and massless spin two. When $2p=d-1$\,,
we end up with only one mode,
\be
	\parbox{42pt}{\begin{tikzpicture}
		\draw (0,0) rectangle (0.8,1.6);
	\draw  (0.4,0) -- (0.4,1.6);
	\draw [dotted] (0,0) rectangle (0.8,-0.4);
	\draw [dotted] (0.4,0) rectangle (0.4,-0.4);
		\node at (-0.3,0.8){${\st p-1}$};
	\end{tikzpicture}}{\phantom{\Bigg|}}^{\sst SO(d-2)}_h
\ee
which we have dualized.
Notice that when $p=1$, this action describes
a scalar mode in three-dimension. 
This is nothing but the propagating content of 
the 3d parity-invariant linear Weyl  gravity
--- or massless limit of New Massive Gravity \cite{Bergshoeff:2009hq} ---
and the scalar mode corresponds 
to the parity-invariant partially-massless spin two.
It would be also interesting to remark
that in $5d$\,, we get in this way a helicity two mode from the action of the $[2,2]$ window field,
which might provide an alternative theory of Gravity.

\subsection{Massive Deformation and New Massive Gravity}
\label{sec:3.3}

The massive deformation of the Weyl action for the generic two-column
fields follows the same pattern as the simple hook case:
\ba
	\cS^{\sst\rm massive}_{\sst [p,q-1]}[\phi,m^2]
	&=& 	\cS^{\sst\L}_{\rm\sst W}[\phi]+m^2\,\cS^{\sst\L}_{\sst{\rm E}\,[p,q-1]}[\phi] \nn
	&\simeq&
	\left(m^2_\L+m^2\right)\left(\la h\,|\,\cG^{\sst \L}_{\sst [p-1,q]}\,h\ra
	-\la f\,|\left(\cG^{\sst \L}_{\sst [p,q-1]}-m^2\,\cI\right) f\ra\right),\\
	\cS^{\sst\rm massive}_{\sst [p-1,q]}[\phi,m^2]
	&=& 	\cS^{\sst\L}_{\rm\sst W}[\phi]+m^2\,\cS^{\sst\L}_{\sst{\rm E}\,[p-1,q]}[\phi]\nn
	&\simeq& 
	\left(m^2_\L-m^2\right)\left(\la h\,|\left(\cG^{\sst \L}_{\sst [p-1,q]}-m^2\,\cI\right) h\ra
	-\la f\,|\,\cG^{\sst \L}_{\sst [p,q-1]}\,f\ra\right),
\ea
which describe one massive $[p,q]$ mode
and one massless $[p,q]\oplus [p,q-1]$ or $[p,q]\oplus[p-1,q]$ mode, where the latter massless mode
becomes partially-massless for $q=p$ case.

Let us consider the dimensions $d=p+q+1$\,,
where the leading $[p,q]$-helicity modes disappear leaving only
lower helicity modes. 
In this case, there is a preferred choice of massive deformation
between $\cS^{\sst\rm massive}_{\sst [p,q-1]}$
and $\cS^{\sst\rm massive}_{\sst [p-1,q]}$ depending on the sign
of the cosmological constant as in the simple hook case.
In dS background, only the massive action $\cS^{\sst\rm massive}_{\sst [p,q-1]}$
can describe massive and massless modes
with the same sign of kinetic terms because
\be
	\cS^{\sst\rm massive}_{\sst [p,q-1]}\sim (m^2_\L-m^2)
\left(\,
\parbox{22pt}{\begin{tikzpicture}
	\draw (0,0) rectangle (0.8,1.6);
	\draw (0.4,0) -- (0.4,1.6);
	\draw [dotted] (0,0) rectangle (0.4,-0.8);
	\node at (0.2,0.8){${\st q}$};
	\node at (0.6,0.8){${\st q}$};
	\end{tikzpicture}}{\phantom{\Bigg|}}^{\sst SO(d-\bm 1)}_h
	+\,\L\ 
	\parbox{22pt}{\begin{tikzpicture}
	\draw (0,0) rectangle (0.8,1.6);
	\draw (0.4,0) -- (0.4,1.6);
	\draw [dotted] (0,0) rectangle (0.4,-0.8);
	\node at (0.2,0.8){${\st q}$};
	\node at (0.6,0.8){${\st q}$};
	\end{tikzpicture}}{\phantom{\Bigg|}}^{\sst SO(d-\bm 2)}_f
	\right).
\ee
Instead in AdS background, the other action 
$\cS^{\sst\rm massive}_{\sst [p-1,q]}$ is preferred since
\be
	\cS^{\sst\rm massive}_{\sst [p-1,q]}\sim (m^2_\L+m^2)
\left(\,\L\,
\parbox{42pt}{\begin{tikzpicture}
	\draw (0,0) rectangle (0.8,1.2);
	\draw (0.4,0) -- (0.4,1.2);
	\draw [dotted] (0,0) rectangle (0.4,-1.2);
	\draw [dotted] (0.4,0) rectangle (0.8,-0.4);
	\node at (-0.3,0.6){${\st q-1}$};
	\end{tikzpicture}}{\phantom{\Bigg|}}^{\sst SO(d-\bm 2)}_h 
	-\, 
	\parbox{22pt}{\begin{tikzpicture}
	\draw (0,0) rectangle (0.8,1.6);
	\draw (0.4,0) -- (0.4,1.6);
	\draw [dotted] (0,0) rectangle (0.4,-0.8);
	\node at (0.2,0.8){${\st q}$};
	\node at (0.6,0.8){${\st q}$};
	\end{tikzpicture}}{\phantom{\Bigg|}}^{\sst SO(d-\bm 1)}_f
	\right).
\ee
Notice that in the above formulas two different Young diagrams
--- one carrying  $SO(d-2)$ representation
and the other carrying $SO(d-1)$ representation --- are used.

Now taking the long-window case ($2p=d-1$), we have two analogue massive theories,
\be 
	\cS^{\sst\rm massive}_{\sst [p,p-1]}\sim
	\parbox{22pt}{\begin{tikzpicture}
	\draw (0,0) rectangle (0.8,2);
	\draw  (0.4,0) -- (0.4,2);
	\node at (0.2,1){${\st p}$};
	\node at (0.6,1){${\st p}$};
	\end{tikzpicture}}{\phantom{\Bigg|}}^{\sst SO(d-\bm 1)}_h\,,
	\qquad
	\cS^{\sst\rm massive}_{\sst [p-1,p]}
	\sim
	\left(\,\L\,
	\parbox{42pt}{\begin{tikzpicture}
		\draw (0,0) rectangle (0.8,1.6);
	\draw  (0.4,0) -- (0.4,1.6);
	\draw [dotted] (0,0) rectangle (0.8,-0.4);
	\draw [dotted] (0.4,0) rectangle (0.4,-0.4);
		\node at (-0.3,0.8){${\st p-1}$};
	\end{tikzpicture}}{\phantom{\Bigg|}}^{\sst SO(d-\bm 2)}_h
	-
	\parbox{22pt}{\begin{tikzpicture}
	\draw (0,0) rectangle (0.8,2);
	\draw  (0.4,0) -- (0.4,2);
	\node at (0.2,1){${\st p}$};
	\node at (0.6,1){${\st p}$};
	\end{tikzpicture}}{\phantom{\Bigg|}}^{\sst SO(d-\bm 1)}_f \right).
\ee
Hence, the massive action $\cS^{\sst\rm massive}_{\sst [p-1,p]}$
having the `partially-massless' gauge symmetry 
have one additional $[p-1,p-1]$ helicity mode
compared to the action $\cS^{\sst\rm massive}_{\sst [p,p-1]}$
corresponding to `massless' gauge symmetry.
The action $\cS^{\sst\rm massive}_{\sst [p,p-1]}$
has only single massive irreducible $[p,p]$ mode, hence
unitary in both of AdS$_{2p+1}$ and dS$_{2p+1}$ background.
On the contrary, the other action $\cS^{\sst\rm massive}_{\sst [p-1,p]}$
propagates two modes with the same kinetic term sign
only in AdS background.
Focusing on the $p=1$ case,
the action $\cS^{\sst\rm massive}_{\sst [1,0]}$
describes a massive spin two in three dimensions and actually
coincides with the linearization of the New Massive Gravity \cite{Bergshoeff:2009hq}.
We can actually see here, that in AdS$_3$ there is an alternative ``New Massive Gravity'', that makes use of the partially-massless symmetry instead of the diffeomorphism one, and contains additional scalar in the spectrum.

As already discussed in previous sections, the term ``New Massive Gravity'' in higher dimensions refers to the models with fields of type $[p,1]$ (see \cite{Joung:2012sa,Dalmazi:2012dq} for related discussion) in dimensions $d=p+2$.
Let us note, that the flat limit of (A)dS$_{p+2}$ New Massive Gravity in dimensions higher than three is not smooth. In flat limit we have only a massive spin two mode, while in AdS we have unitary model with (massive spin two + massless scalar), and in dS the unitary model contains (massive spin two + massless spin two).
Only in three dimensions, one can have New Massive Gravity with the same spectrum around flat and constantly curved backgrounds, at least at the linearized level.


\acknowledgments

We thank Dario Francia
for discussions which incited our interest in the problem explored here.
The work of EJ was supported in part by the National Research Foundation of Korea through the grant NRF-2014R1A6A3A04056670 and the Russian Science Foundation grant 14-42-00047 associated with Lebedev Institute. 
 The work of KM was supported by the BK21 Plus Program funded by the Ministry of Education (MOE, Korea) and National Research Foundation of Korea (NRF).

\bibliographystyle{JHEP}
\bibliography{Ref}

\end{document}